\documentclass[letterpaper, 10 pt, conference]{ieeeconf}  

\IEEEoverridecommandlockouts                              
\usepackage{graphicx} 
\usepackage{bm}
\usepackage{datetime}
\usepackage{multicol}

\title{\LARGE \textbf{Omnichain Web}\\
\large The Universal Framework for Streamlined Chain Abstraction and Cross-Layer Interaction
}

\author{Hardik Gajera \qquad \qquad \qquad \qquad \quad Akhil Reddy \quad \qquad \qquad \qquad \qquad Bhagath Reddy\\hardik@dojima.foundation\qquad \qquad akhil@dojima.foundation \qquad \qquad bhagath@dojima.foundation \\
\\\vspace*{20pt} \normalsize{November 2024}
}

\begin{document}

\thispagestyle{empty}
\pagestyle{empty}

\twocolumn[
\begin{@twocolumnfalse}
\maketitle
\begin{abstract}
The Web3 ecosystem is highly fragmented, making it difficult for over a billion Web2 businesses, traditional enterprises, and thousands of AI protocols to integrate seamlessly. With the rapid growth of blockchains, rollups, and app-specific chains, cross-chain interactions remain inefficient, and liquidity is deeply fragmented. AI systems, which thrive on data, lack standardized access to blockchain ecosystems, limiting their ability to function autonomously. Furthermore, intent-based interactions, crucial for AI-driven automation, are constrained by the absence of scalable platforms that enable users and developers to express and execute intents across chains. The current solver ecosystem is centralized, with most of the volume controlled by project teams, due to the lack of developer-friendly tools that address liquidity fragmentation and rebalancing challenges.

To solve this multilayered, multidimensional challenge, Dojima's Omnichain Web introduces a universal framework that abstracts blockchain complexity and unifies Web2, Web3 and AI systems. OmniRollups, Proof Network, Ragno Network, and Builder Marketplace form its core pillars. The Omni Sequencer ensures atomic, secure execution across chains, while Linera microchains enable AI-driven transaction automation. Ragno Network serves as a decentralized marketplace for L1 infrastructure, while the Proof Network provides cryptographic security for omnichain transactions. The Builder Marketplace introduces a solver-driven execution layer, empowering developers to build and monetize intent-based applications without liquidity constraints. By enabling a composable marketplace at the crossroads of Web2 and Web3, Omnichain Web fosters a seamless flow of data, value, and computation across previously isolated ecosystems.

Omnichain Web is a universal framework that mirrors the internet that unifies Web3 and Web2, enabling a seamless, composable application marketplace. By combining Web2 scale with Web3 decentralization, we are building the foundation for the next wave of adoption. 
\end{abstract}
\vspace*{30pt}
\end{@twocolumnfalse}
]

\section{\textbf{INTRODUCTION}}
\vspace*{10pt}
The evolution of the Web3 space has mirrored the development of national economies and their financial systems. Just as countries have their own currencies and central banks, blockchain networks have their own tokens and foundations. Understanding these parallels provides insights into the decentralized world of cryptocurrencies and blockchain technology, highlighting the critical need for interoperability and robust settlement frameworks~\cite{crosschain}.

\begin{figure}[h!]
    \centering
    \includegraphics[width=8cm, height=7cm]{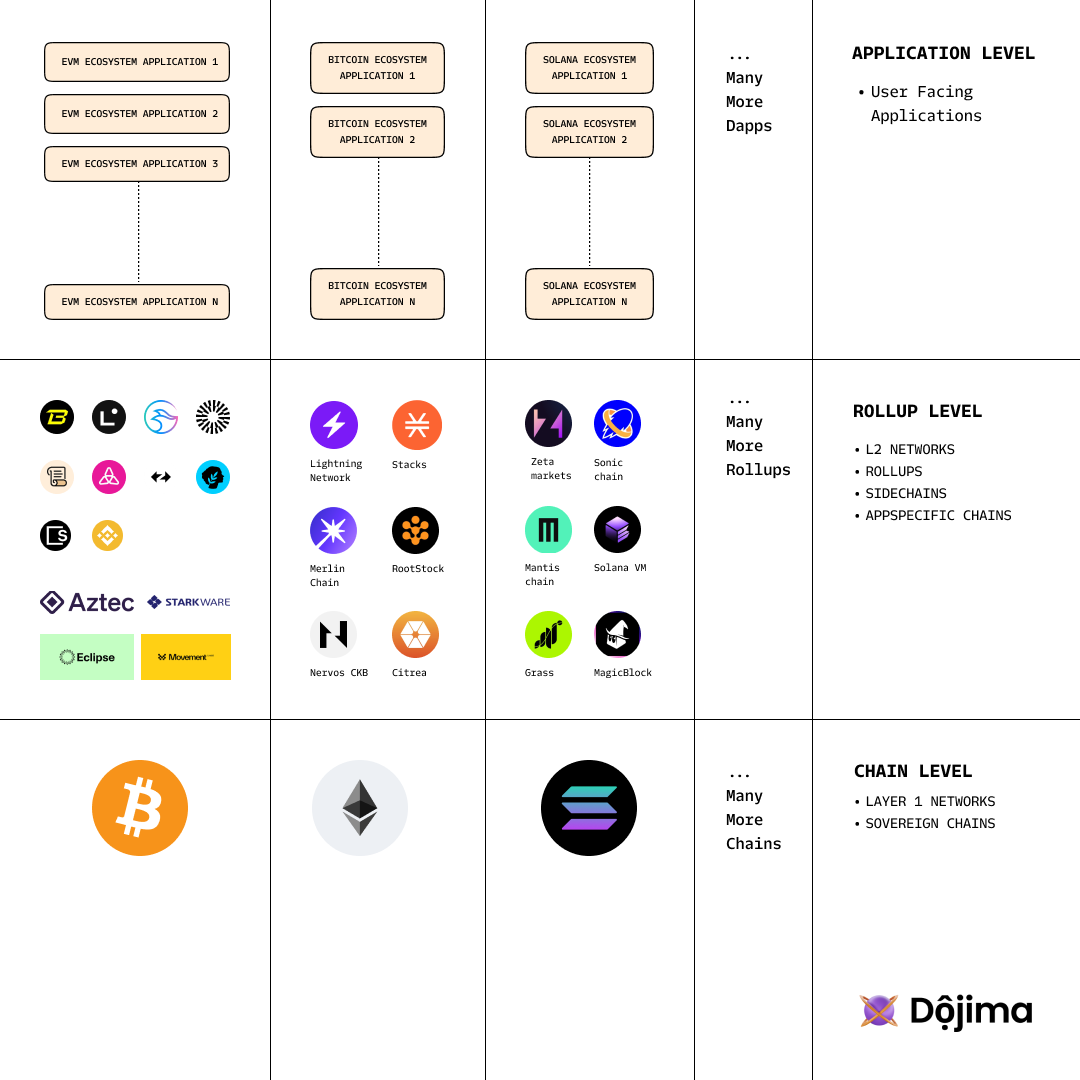}
    \caption{Fragmented Space}
    \label{fig:fragmented}
\end{figure}

\begin{itemize}
    \item {Global Financial Messaging}: The Society for Worldwide Interbank Financial Telecommunication (SWIFT) provides a secure and standardized messaging network for international payments between financial institutions. In a decentralized world, an equivalent to SWIFT is essential for enabling seamless cross-chain communication and interoperability. Protocols like Chainlink~\cite{chainlink} and Cosmos~\cite{cosmos} aim to bridge different blockchains, facilitating secure and efficient data and asset transfers.
    \item {Interoperability and Standardization:} Just as SWIFT standardizes financial messages, decentralized interoperability solutions standardize data formats and communication protocols between blockchains. This standardization is crucial for the efficient functioning of the multi-chain ecosystem.
\end{itemize}

\begin{figure}
    \centering
    \includegraphics[width=7cm, height=7cm]{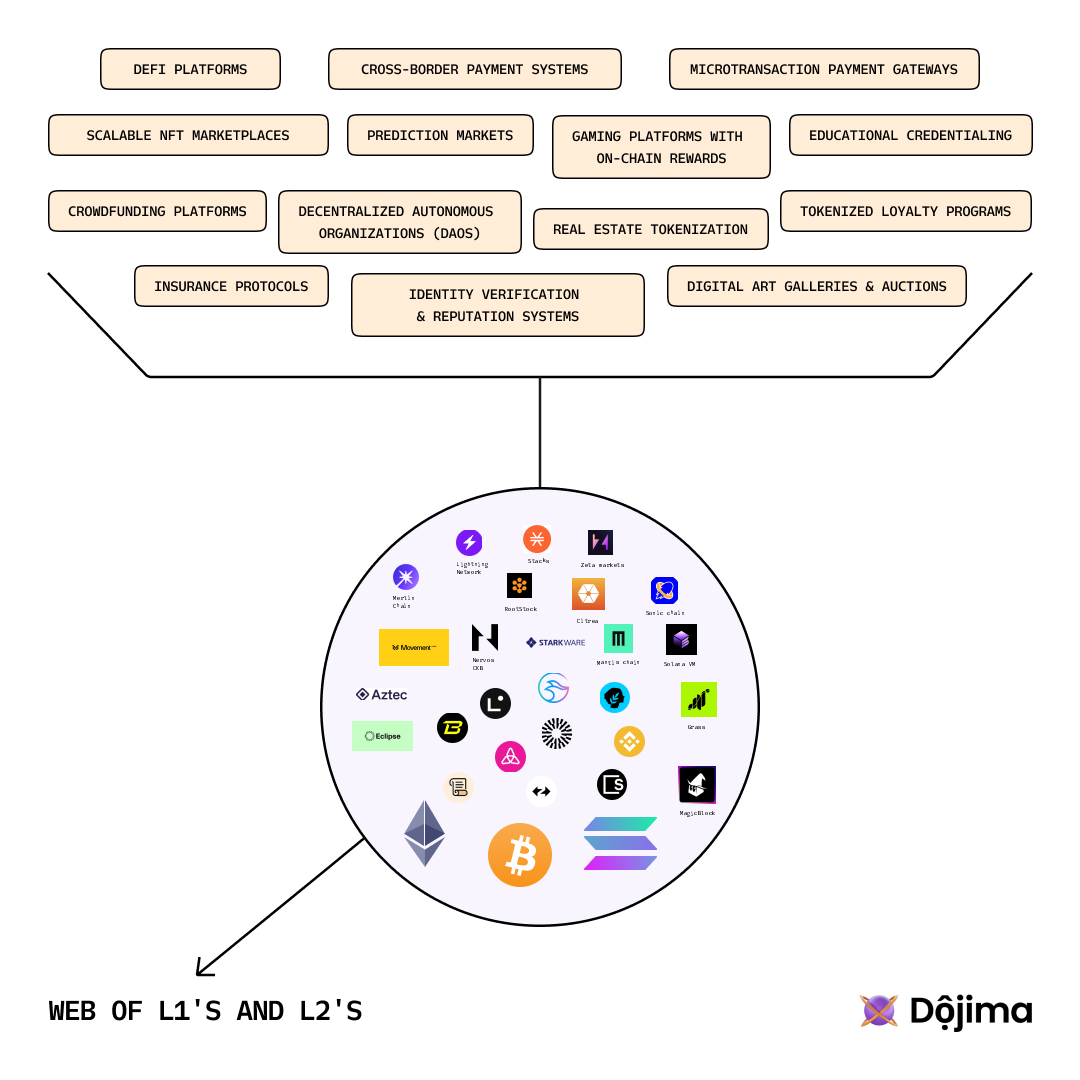}
    \caption{Omniweb}
    \label{fig:omniweb}
\end{figure}
The comparison of countries to blockchains and central banks to blockchain foundations highlights the structured yet diverse nature of the Web3 ecosystem. Just as countries require robust financial systems to thrive, blockchain networks need effective governance and interoperability solutions. In this decentralized world, networks like SWIFT and cross-chain settlement mechanisms play a pivotal role in ensuring seamless and secure interactions between different blockchains and rollups. As the Web3 space continues to evolve, these foundational elements will be crucial in driving innovation and achieving a truly interconnected digital economy.

The current cross-chain ecosystem faces several key challenges, including:
\begin{itemize}
    \item \textbf{Liquidity Fragmentation:} When solvers execute user intents across various chains, they receive repayments on the originating chains, resulting in capital being scattered across numerous Layer 1 (L1) and Layer 2 (L2) networks. This leads to inefficiencies, as significant portions of liquidity become inactive or "dead weight." Thinly spread liquidity diminishes the ability of solvers to effectively manage and respond to cross-chain demands, increasing the operational complexity and costs.\\

    \item \textbf{Lack of Unified Capital Flow Management:} There is no shared system for coordinating capital flows between different chains, forcing solvers to manage liquidity independently. Each solver operates in isolation, rebalancing funds manually across multiple networks, which is time-consuming and costly. This leads to duplicated efforts and fragmented liquidity management, causing high operational expenses and limiting the scalability of cross-chain activities.\\

    \item \textbf{Lack of Platforms for User-Centric Applications:}
    Current cross-chain solutions are heavily infrastructure-focused, lacking an ecosystem that supports the development of user-centric decentralized applications (dApps). The fragmented state of blockchain interoperability makes it difficult for developers to create seamless user experiences across multiple chains. This limits the adoption of Web3 technologies by everyday users, as navigating multiple networks and wallets becomes cumbersome. A unified platform is needed to enable developers to build applications that offer smooth, cross-chain user experiences.\\

    \item \textbf{Limited Compatibility with AI Agents:}
    The AI market is maturing, with players like ChatGPT and Perplexity readily available to every PC and mobile user. Recently, OpenAI announced that they have begun working on an AGI model. In this context, within the next 6-12 months, we are likely to see AI agents working and acting on behalf of users, including executing transactions.~\cite{aiagent}. 

    As AI agents become more integrated into Web3 solutions, there is a lack of infrastructure that supports their seamless operation across diverse blockchain networks. Existing systems do not provide robust mechanisms for intent-based interactions or automation powered by AI, which can dynamically interact with various chains based on user requirements. This gap hinders the deployment of advanced AI-driven functionalities, such as automated asset management or predictive trading strategies, within decentralized ecosystems. A platform that supports compatibility with AI agents would enhance the ability to create smart, adaptive dApps capable of complex decision-making.
\end{itemize}

Everclear~\cite{everclear} addresses liquidity fragmentation issue by introducing a "Clearing Layer" for global netting and settlement of capital flows. It functions primarily across Layer 2 (L2) rollups and higher-level chains, coordinating liquidity rebalancing through a decentralized network built on Arbitrum Orbit and utilizing EigenDA for data availability. By aggregating and netting transactions across chains, Everclear can significantly lower the cost of liquidity transfers, offering periodic rebalancing (approximately every 3-6 hours), which helps solvers maintain optimal liquidity levels without constant manual intervention. However, its focus is mainly on L2 chains and selected assets, operating with certain limitations like a permissioned list of supported assets and chains. It also relies on Eigenlayer for security, which currently lacks slashing mechanisms to enforce strong economic security~\cite{lifi}. These constraints limit Everclear's applicability in a fully decentralized and permissionless context.

Omnichain Web extends beyond these limitations by offering true cross-layer interoperability, leveraging OmniRollups—a highly interoperable rollup solution that enables seamless asset settlement across both Layer 1 (L1) and Layer 2 (L2) chains. Through an adapter, it can also connect existing rollups, transforming them into OmniRollups that enable cross-chain asset settlement across multiple L1s. This allows existing rollups to integrate with Omnichain Web’s ecosystem without needing to overhaul their existing infrastructure. The use of OmniRollups and its compatibility adapter create a broader, more decentralized network for asset settlement, bridging both L1 and L2 chains. This expanded functionality offers solvers and network participants a unified liquidity management and cross-chain transaction system without the limitations of a centralized clearing mechanism. Consequently, Omnichain Web provides a robust and scalable solution for fragmented liquidity management, delivering efficiency gains and enhancing the usability of cross-chain protocols.

Moreover, to enable seamless compatibility with AI agents, Omnichain Web is developing specialized endpoints and adapters tailored for various AI models. These components will allow AI agents to interact effectively with the decentralized ecosystem. By leveraging a Trusted Execution Environment (TEE), Omnichain Web will provide a secure layer for handling sensitive transactions and executing operations autonomously. This secure setup ensures that AI agents can dynamically interact with the abstracted cross-chain world built by Omnichain Web, facilitating real-time decision-making and intent execution across multiple blockchains without compromising security or efficiency. 
    
\vspace*{10pt}

\section{\textbf{Omnichain Web ecosystem}}
\vspace*{10pt}

\begin{figure}
    \centering
    \includegraphics[width=6cm, height=17cm]{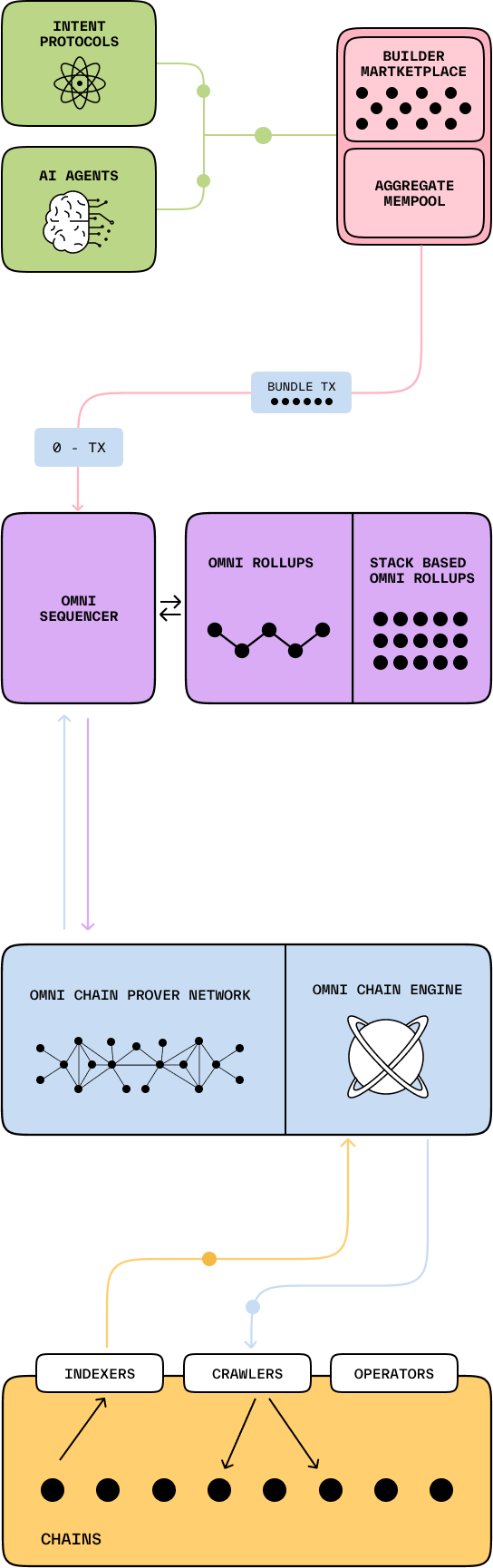}
    \caption{Omnichain Web Architecture}
    \label{fig:omni-overview}
\end{figure}
The Omnichain Web is a comprehensive framework designed to address the evolving needs of the multichain blockchain ecosystem. As blockchain applications grow more complex and span across multiple Layer 1 (L1) and Layer 2 (L2) networks, the need for seamless cross-chain interoperability, secure verification mechanisms, and efficient data aggregation has become critical. As described in the Figure~\ref{fig:omni-overview}, the Omnichain Web integrates these elements through its four core components: the Builder Marketplace, Omni Rollups, Proof Network, and Ragno Network. Each part plays a distinct role in ensuring that developers and users can interact with decentralized applications across multiple blockchains with reduced friction and enhanced security.

It all starts with the Ragno Network, a decentralized marketplace weaving L1's together. It offers unified access points, making it easier to move assets across chains and supporting cross-chain protocols and omnichain applications. Its role is to facilitate seamless cross-chain interactions by providing a reliable infrastructure for indexing and querying data across multiple blockchains. The Ragno Network indexes transaction details and state changes, making it possible to retrieve and verify information across various chains efficiently. However, connectivity between different chains need trust. That’s where the Proof Network and Omnichain Engine come in. The Ragno network receives block details and proofs from the Proof Network, which are then used to update its indexing services, ensuring that all interactions are accurately reflected across the omnichain ecosystem. The zk-proving layer acts like a shield, enabling verifiable, secure transactions across chains and ensuring reliable proof distribution.

This foundation supports the Omnichain rollups, transforming traditional L2's into omnichain rollups. Traditional rollup solutions focus on a single chain, which limits their ability to interact with other networks. Omni Rollups extend this capability by converting multiple isolated rollup stacks into a cohesive omnichain rollup layer. This architecture allows for multichain settlement, where transactions can be initiated on one chain and settled across different chains seamlessly. By leveraging the Omni Rollups, developers can build applications that natively support cross-chain functionalities, such as transferring assets or data between different blockchains, thereby enabling a truly omnichain experience for users. The Builder Marketplace completes the ecosystem by solving intents without relying on imperative chains, making it easier for developers to create custom omnichain solutions.

Imagine a user wants to swap assets between Ethereum (L1) and a Layer 2 network like Arbitrum. The dApp developer initiates the swap request through the Builder Marketplace, which bundles this request and sends it to the Omni Rollups. The sequencer processes the swap, executes the transaction on both Ethereum and Arbitrum, and generates a proof via the Proof Network. The validated proof is then passed to the Ragno Network, where the cross-chain swap is settled, and the updated balances are indexed and made available for querying. This end-to-end workflow is performed in a matter of seconds, providing a seamless experience for the user without manual intervention across chains.

\vspace*{10pt}

\subsection{\textbf{Builder Marketplace}}
\vspace*{10pt}
The Builder Marketplace in the Omnichain Web architecture acts as the topmost layer, serving as a gateway for developers to access decentralized services across multiple blockchains. It integrates an innovative approach by utilizing Intent Protocols and AI Agents, which are deployed within a Trusted Execution Environment (TEE) for enhanced security and reliability. 

Intent Protocols are specialized frameworks that allow users to express high-level intents or actions they wish to perform on decentralized applications, rather than specifying low-level transaction details. For example, instead of manually transferring assets between chains, a user can simply state their intent to swap tokens across networks. The Intent Protocol then translates this high-level request into actionable steps, selecting the optimal chain or protocol for execution based on factors like cost, speed, and liquidity availability. This abstraction simplifies user interactions and reduces the complexity involved in cross-chain transactions. AI Agents within the Builder Marketplace play a crucial role in processing and optimizing these intents. These agents leverage machine learning models to predict the most efficient execution path for a given intent. For instance, they can analyze network conditions, gas fees, and liquidity availability to decide which chain or rollup stack to route the transaction through. 

\begin{figure}[ht]
    \centering
    \fbox{\includegraphics[width=7cm, height=2.5cm]{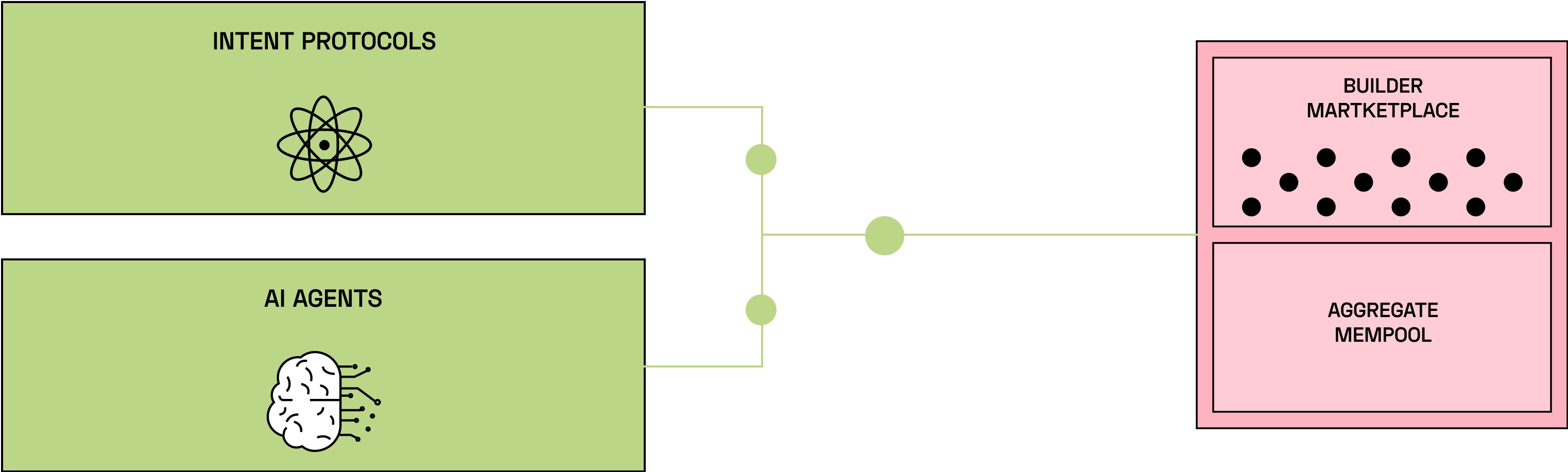}}
    \caption{Builder Marketplace}
    \label{fig:builder}
\end{figure}

Deploying intent protocols and AI agents within a TEE ensures that they operate in a secure and isolated setting, safeguarding the integrity and confidentiality of both the input data and the computational process. The Builder Marketplace facilitates this by offering specialized endpoints and adapters tailored for AI models, intent protocols, and even traditional web2 applications. These components enable diverse apps to seamlessly interact with the decentralized ecosystem of Omnichain Web. The TEE's secure execution not only processes sensitive data without exposure to external threats but also generates an attestation report alongside the output~\cite{phala}. This cryptographic proof verifies that the transaction was conducted in a secure, tamper-proof environment, assuring participants of its trustworthiness and integrity. This setup bridges the gap between AI agents, web2 apps, and decentralized protocols, creating a unified, secure platform for intent execution across multiple chains.

\vspace*{5pt}
\noindent
\textbf{Support for Multiple TEE Systems}
\vspace*{5pt}

To cater to a diverse set of developers and programming languages, the Builder Marketplace supports multiple TEE systems. This flexibility allows developers to choose their preferred language and platform:
\begin{itemize}
    \item Rust-based TEE using Teaclave TrustZone SDK~\cite{teaclave}: For Rust developers, the system utilizes the Teaclave TrustZone SDK (Rust OP-TEE TrustZone SDK), enabling the creation of secure TrustZone applications. The Teaclave framework, deployed on a QEMUv8 platform, offers a safe and efficient environment for building and running Rust applications, leveraging the capabilities of ARM TrustZone technology.
    \item JavaScript-based TEE using WASM and Intel SGX~\cite{phala}: For JavaScript developers, the system integrates a JavaScript engine within a WASM-based Virtual Machine (VM), running over Intel's Software Guard Extensions (SGX), similar to the approach used by the Phala Network. This setup allows developers to write secure, TEE-based JavaScript applications, leveraging the robustness of Intel SGX for secure execution.
\end{itemize}

\vspace*{5pt}
\noindent
\textbf{Mempool Aggregator and Transaction Bundling}
\vspace*{5pt}

The attestation report and the output generated by the TEE are passed to the Builder Marketplace, where a Mempool Aggregator verifies the attestation report. The Mempool Aggregator ensures that the transactions have been processed securely within a TEE, validating the integrity of the execution before bundling the transactions. Once verified, these transactions are aggregated into bundles and sent to the Omni Rollups layer for further processing. By integrating intent protocols, AI-driven optimization, and TEE-based secure execution, the Builder Marketplace provides a robust, scalable, and secure entry point for decentralized applications across multiple chains. 

This setup not only enhances the security of transaction processing but also offers a flexible and developer-friendly environment by supporting multiple languages and TEE systems, empowering a broader range of developers to build omnichain solutions efficiently.

\vspace*{10pt}
\subsection{\textbf{Omni Rollups}}
\vspace*{10pt}
The Omni Rollups layer in the Omnichain Web architecture is a powerful component that transforms traditional rollup capabilities into a highly interoperable Omnichain Rollup ecosystem. Leveraging innovations introduced by Dojima Network V2, this layer enables seamless interactions with multiple Layer 1 (L1) blockchains, significantly enhancing the cross-chain functionality and liquidity in the Web3 space.

\begin{figure}[ht]
    \centering
    \fbox{\includegraphics[width=8cm, height=1.8cm]{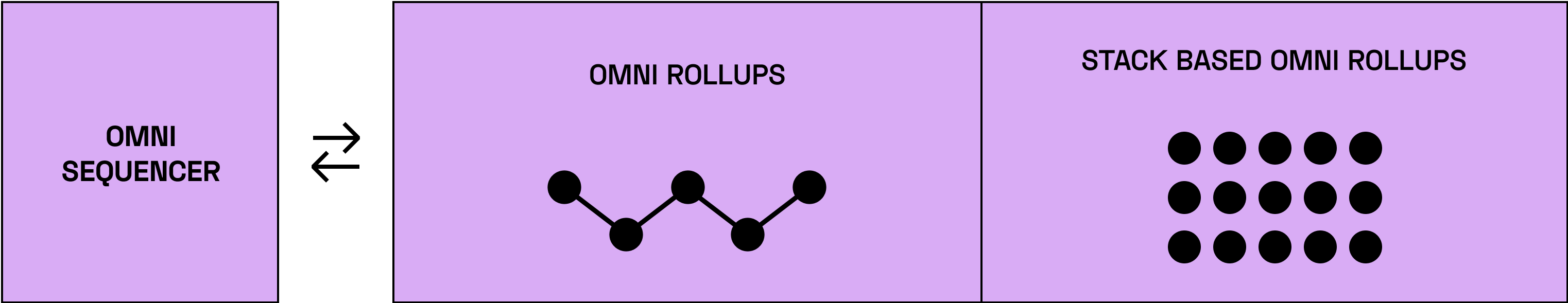}}
    \caption{Omnirollups}
    \label{fig:omnirollups}
\end{figure}

Omni Rollups bring a novel approach to rollup technology by allowing rollup stacks to natively interact with multiple L1s, rather than being confined to a single chain. This eliminates the typical one-to-one dependency model seen in traditional rollups, unlocking one-to-many interactions between rollups and L1 blockchains. As a result, Omni Rollups can directly fetch and settle assets across various L1s without relying on a single base layer. This provides several key benefits:

\begin{itemize}
    \item Direct Asset Fetching and Settlement: Omni Rollups can fetch assets directly from integrated L1 networks and return them to their respective source chains. This capability ensures that users can access liquidity and perform transactions across different blockchains without the need for intermediate layers or centralized bridges, thus reducing complexity and enhancing the user experience.
    \item Faster Confirmations and Settlements: The enhanced interoperability of Omni Rollups allows for quicker confirmations and settlements by directly interacting with multiple L1 chains. This reduces the latency typically associated with cross-chain transactions and improves the speed of asset transfers in a multichain environment.
    \item Increased Liquidity and Asset Availability: By enabling seamless interactions with multiple L1s, Omni Rollups enhance cross-chain asset liquidity. Users can easily move assets across different chains, and developers can build dApps that leverage the liquidity of various L1s without being constrained to a single chain.
\end{itemize}

\vspace*{5pt}
\noindent
\textbf{Enhanced Interactions with L1 Blockchains}
\vspace*{5pt}

The Omni Rollups layer significantly benefits existing rollup stacks by enhancing their ability to interact with various L1 blockchains in a flexible and efficient manner:

\begin{itemize}
    \item One-to-Many Interactions: Unlike traditional rollups, which typically interact with a single L1, Omni Rollups support one-to-many interactions. This means a single rollup stack can simultaneously communicate with multiple L1 networks, allowing dApps to execute complex cross-chain operations, such as atomic swaps and multichain liquidity pooling, with minimal friction.
    \item Improved Settlement Speed: By directly settling assets with the respective L1s, Omni Rollups reduce the time required for finalizing transactions. This enhances the overall user experience by providing faster and more reliable settlement times, which is particularly valuable for applications requiring real-time interactions, such as DeFi protocols and decentralized exchanges.
\end{itemize}

 Omni Rollups serve as the pivotal layer that bridges isolated rollup stacks and diverse L1 networks, creating a unified, interoperable ecosystem. By offering direct asset fetching, faster settlement, and enhanced liquidity, Omni Rollups provide developers and users with a scalable and efficient platform for building and interacting with decentralized applications across multiple blockchains. This innovation redefines the capabilities of rollups, transforming them from isolated scaling solutions into integral components of the omnichain infrastructure, driving forward the vision of a truly interoperable Web3.
\vspace*{10pt}
\subsection{\textbf{Proof Network}}
\vspace*{10pt}
The Proof Network is designed to support a wide range of proof systems, enabling comprehensive verification across various Layer 1 (L1) and Layer 2 (L2) blockchains. This network plays a crucial role in maintaining the integrity and trustworthiness of transactions as they move between different rollups and chains. Aligned Layer~\cite{aligned} and Unichain~\cite{unichain} focus on providing interoperability solutions and secure multi-chain execution environments. However, while Aligned Layer emphasizes a protocol-agnostic, cross-layer approach with an integrated data layer, Unichain is centered on building a unified smart contract platform to streamline interactions across different chains. In contrast, Proof Network supports a modular zkVM-based proof system with recursive proofs to verify and secure cross-chain transactions, focusing on ensuring trust and validation in a decentralized manner across both L1 and L2 chains.

\begin{figure}[ht]
    \centering
    \fbox{\includegraphics[width=8cm, height=1.8cm]{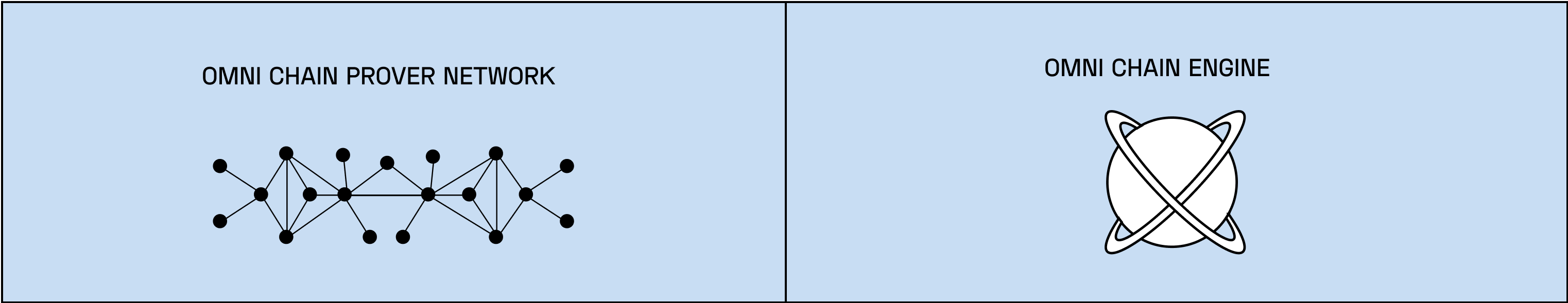}}
    \caption{Proof Network}
    \label{fig:proofnetwork}
\end{figure}

At the core of the Proof Network is a modular zkVM (Zero-Knowledge Virtual Machine), which is specifically designed to handle multiple proof systems. This flexibility allows developers to choose from various zero-knowledge proof (ZKP) techniques, such as Groth16~\cite{groth}, zk-STARKs~\cite{zkstark}, Plonk~\cite{plonk}, Hyperplonk~\cite{hyperplonk}, and Binius~\cite{binius}, based on the specific requirements of their decentralized applications (dApps) or protocols. The modular design of the zkVM ensures that it can seamlessly integrate new and emerging proof systems, making the Proof Network future-proof and adaptable to the evolving needs of the blockchain ecosystem.

To enhance efficiency and reduce the computational overhead, the Proof Network employs a recursive proof system. This advanced technique allows it to aggregate multiple proofs from different transactions or rollups into a single, unified proof. By using recursive proofs, the network can verify a batch of diverse proofs in a single computation, significantly optimizing the verification process. This reduces the time and gas costs associated with verifying multiple transactions across various chains, making the cross-chain interactions faster and more scalable.
\vspace*{10pt}
\subsection{\textbf{Ragno Network}}
\vspace*{10pt}
The Ragno Network offers a shared, decentralized infrastructure for cross-chain protocols and light clients, eliminating the need for each validator to run separate nodes for every L1 blockchain. Instead, Ragno provides a pre-existing, decentralized L1 infrastructure that can be leveraged by validators, light clients, and dApps.

\begin{figure}[ht]
    \centering
    \fbox{\includegraphics[width=8cm, height=1.8cm]{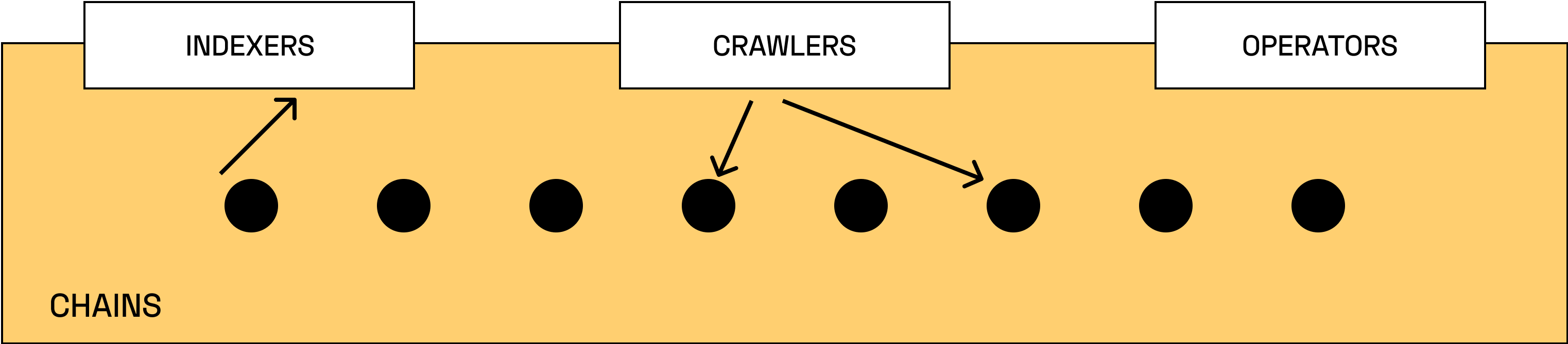}}
    \caption{Ragno Network}
    \label{fig:ragno}
\end{figure}

\vspace*{5pt}
\noindent
\textbf{Key Benefits of Ragno Network}
\vspace*{5pt}

\begin{itemize}
    \item Cost Savings: By utilizing Ragno's shared infrastructure, validators can reduce their operating costs multiple fold. This removes redundant node setups, significantly lowering the barrier to entry for cross-chain participation.
    \item Enhanced Efficiency: Validators and participants can access reliable cross-chain data through Ragno's network without the overhead of managing multiple L1 nodes. This streamlined approach increases overall efficiency and reduces operational complexities.
    \item Decentralization and Security: Built on robust cryptographic principles, Ragno Network prioritizes a high degree of decentralization and security. Its architecture ensures trustworthy and tamper-proof cross-chain data interactions, essential for maintaining the integrity of blockchain transactions.
    \item Modular Infrastructure: The Ragno Network offers a modular approach by decoupling the data layer from the execution layer, providing:
    \begin{itemize}
        \item Flexible Data Indexing: Optimized data indexing tailored to the specific requirements of different blockchain integrations, enhancing performance.
        \item Faster Transaction Mapping: Efficient transaction mapping reduces costs and speeds up the integration process with new blockchains, facilitating a smoother user experience and quicker protocol upgrades.
    \end{itemize}
         
\end{itemize}

With its vision to streamline cross-chain interactions and reduce operational overhead, Ragno Network aims to be the foundational data layer for decentralized applications, driving the next generation of scalable, interoperable blockchain solutions.

\vspace*{10pt}
\section{\textbf{Conclusion}}
\vspace*{10pt}

The future of blockchain lies in unifying diverse chain ecosystems, as relying solely on a single blockchain cannot meet the growing demand for seamless, scalable, and user-centric applications. Instead, the emphasis is shifting towards a cross-chain ecosystem, where interoperability and fluid asset transfers are key to unlocking the full potential of decentralized technologies. Omnichain Web is at the forefront of this movement, creating a comprehensive platform that not only bridges multiple Layer 1 (L1) and Layer 2 (L2) blockchains through its OmniRollups but also integrates traditional web2 applications with web3 protocols by providing speacialized endpoints to link web2 with builder marketplace. By providing compatibility layers for AI agents and enabling a secure environment for executing complex user intents, Omnichain Web is pioneering a holistic, user-centered framework that brings together web2 efficiency and web3 innovation, paving the way for the next generation of decentralized applications. Omnichain Web is not just a framework; it is the blueprint for the future of decentralized interoperability.

\end{document}